\documentclass[english,prx,twocolumn,superscriptaddress,longbibliography]{revtex4-1} 
\usepackage[bookmarks=false,linkcolor=blue,urlcolor=blue,colorlinks,citecolor=blue]{hyperref} 
\usepackage[T1]{fontenc} 
\usepackage[latin9]{inputenc} 
\usepackage{amsmath} 
\usepackage{graphicx} 
\usepackage{babel} 
\usepackage{color}

\newcommand{\Ket}[1]{\left|#1\right>}

\begin{document}

\title{The Transformation of the Superconducting Gap to an Insulating Pseudogap \\
at a Critical Hole Density in the Cuprates} 
\begin{abstract}
	We apply the recent wavepacket formalism developed by Ossadnik to describe the origin of the short range ordered pseudogap state as the hole doping is lowered through a critical density in cuprates. We argue that the energy gain that drives this precursor state to Mott localization, follows from maximizing umklapp scattering near the Fermi energy. To this end we show how energy gaps driven by umklapp scattering can open on an appropriately chosen surface, as proposed earlier by Yang, Rice and Zhang. The key feature is that the pairing instability includes umklapp scattering, leading to an energy gap not only in the single particle spectrum but also in the pair spectrum. As a result the superconducting gap at overdoping is turned into an insulating pseudogap, in the antinodal parts of the Fermi surface. 
\end{abstract}

\author{Ye-Hua Liu} \affiliation{Theoretische Physik, ETH Zurich, 8093 Zurich, Switzerland}

\author{Wan-Sheng Wang} \affiliation{National Laboratory of Solid State Microstructures $\&$ School of Physics, Nanjing University, Nanjing, 210093, China} \affiliation{Department of Physics, Ningbo University, Ningbo 315211, China}

\author{Qiang-Hua Wang} \affiliation{National Laboratory of Solid State Microstructures $\&$ School of Physics, Nanjing University, Nanjing, 210093, China} \affiliation{Collaborative Innovation Center of Advanced Microstructures, Nanjing 210093, China}

\author{Fu-Chun Zhang} \affiliation{Collaborative Innovation Center of Advanced Microstructures, Nanjing 210093, China} \affiliation{Kavli Institute for Theoretical Sciences, University of Chinese Academy of Sciences, Beijing 100190, China} 

\author{T. M. Rice} \affiliation{Theoretische Physik, ETH Zurich, 8093 Zurich, Switzerland} \affiliation{Condensed Matter Physics and Material Science Department, Brookhaven National Laboratory, Upton, New York 11973, USA}

\pacs{xxx}

\maketitle

\section{Introduction} The novel nature of the transition from a full Fermi surface metal into the pseudogap state with a truncated Fermi surface as the doped hole density is reduced, is one of the most studied features of the cuprate superconductors. The interpretation of ARPES experiments \cite{Yang2011} that it is a transition from a conventional Bloch band metal to a precursor state of the Mott insulator at zero doping, has been confirmed by recent high-field Hall effect measurements by Badoux et al \cite{Badoux2016} and Lalibert\'{e} et al \cite{Laliberte2016}. These experiments illustrate the dramatic change in the $\mathbf{k}$-space Fermi surface that appears without changing the translational symmetry of the underlying lattice. A key challenge is a microscopic description of this instability of the Landau Fermi surface to a $\mathbf{k}$-space reorganization as the hole density decreases and the Mott state is approached. Note the strong contrast to the case of $^3$He which remains a Landau Fermi liquid although the onsite repulsion relative to the kinetic energy is much larger. Standard Fermi surface instabilities show up in a Functional Renormalization Group (FRG) analysis as a divergence in a specific particle-hole or particle-particle channel, which can be analyzed by a symmetry lowering mean field theory. However the lack of symmetry breaking associated with the onset of the pseudogap rules out this approach. Our aim is to obtain a microscopic description of the novel instability at the onset of the pseudogap.

The starting point of any instability analysis is the behavior of the effective low energy Hamiltonian, $H_\mathrm{eff}$, as the instability is approached. The early FRG calculations on the 2D Hubbard model by Honerkamp et al \cite{Honerkamp2001} showed strong growth in umklapp processes as the onsite repulsion increased at underdoping. These processes are especially strong for sets of wavevectors on a square surface which join the antinodal points, enclosing half the Brillouin zone. They labeled this the umklapp surface (US). A special feature of the 2D square lattice near $\frac{1}{2}$-filling is the presence of this US close by to the band structure Fermi surface. Usually elastic U-scattering is present only at isolated points on the Fermi surface and as a result can be safely neglected. But here U-scattering appears on the whole US and is crucial to the low energy physics. Besides strong U-processes, the FRG flow also showed strong enhancements in the $d$-wave superconducting ($d$SC) and the antiferromagnetic (AF) susceptibilities \cite{Honerkamp2001,Honerkamp2002}. The presence of strong U-scattering on the US in such a state, motivated Yang, Rice and Zhang \cite{Yang2006,Rice2012} to propose a simple pairing form for the single particle propagator in the pseudogap phase. Their phenomenological YRZ form places the pairing gap on the US, not the Fermi surface, leads to an enhanced energy gap due to U-scattering processes. The YRZ propagator has been used to successfully interpret both detailed ARPES experiments \cite{Yang2011} and recent Hall effect experiments \cite{Storey2016}. More recent FRG improved analyses \cite{Husemann2009,Xiang2012} has led to detailed forms for $H_\mathrm{eff}$ at low energies in terms of a combination of coexisting $d$-wave particle-particle pairing and AF $(\pi,\pi)$ particle-hole terms, analogous to a parquet diagram approach in a diagrammatic formulation.

There is one example of a Mott insulator phase at weak coupling, amenable to $\mathbf{k}$-space analysis, namely the ``D-Mott'' insulator state in the $\frac{1}{2}$-filled 2-leg Hubbard ladder \cite{Balents1996,Konik2001}. This has a fully truncated Fermi surface caused by strictly short range (SRO) correlations. In this example the D-Mott insulating behavior is driven by the extra umklapp (U) scattering processes present exactly at the $\frac{1}{2}$-filling Fermi surface. Short range order appears in both the $d$SC and the commensurate $(\pi,\pi)$ AF channels. The presence of strong and coupled divergences in both these channels cannot lead to coexisting long range ordering in both channels. Singlet pairing results in a spin gap which is incompatible with long range AF order, which in turn would open a charge gap, incompatible with superconductivity. This incompatibility does not allow the resulting strongly coupled state to be treated by mean field theory plus fluctuations. Rather the D-Mott insulator phase of the $\frac{1}{2}$-filled 2-leg Hubbard ladder \cite{Balents1996,Konik2001} can be viewed as a state where both kinds of SRO coexist, stabilized by the energy gain from the presence of local correlations in both particle-hole and particle-particle channels. Their analysis is limited to the weak coupling limit. Recently numerical DMRG methods have been developed which can be applied to arbitrary strength interactions. As will be discussed in detail below, the numerical results show a continuous transition between the weak coupling and strong coupling (i.e. Mott) limit. The only difference is breakdown of the special SO(8) symmetry applies only at weak coupling but the isolated ground state and finite single particle, charge and spin gaps evolve continuously with increasing interaction strength. One important limitation is the real space character of DMRG which limits detailed comparison of the DMRG and RG results.

The physics of this SRO state in the ladders is the same as that in a Resonating Valence Bond (RVB) state that was proposed very early by Anderson \cite{Anderson1987,Anderson2004} as underlying high temperature superconductivity in the cuprates. The initial RVB proposal started from an undoped Mott insulator and discussed hole doping in a 2D Heisenberg $S=\frac{1}{2}$ model. But the $\frac{1}{2}$-filled 2-leg Hubbard ladder shows that the same physics can appear in 1D at weak and strong coupling. This raises the possibility of realizing RVB physics as an instability of a 2D metal as the hole density is reduced towards the Mott state. An important obstacle is the extension of the 1D SRO ladder state to a 2D SRO state. In the next chapter we will introduce the important advance made by Ossadnik on a microscopic theory of SRO states in a 2D lattice.

\section{The Ossadnik-WW Theory for SRO in interacting fermions}

The challenge to construct a microscopic theory for systems with strictly SRO and no broken symmetries, inspired Matthias Ossadnik \cite{Ossadnik2016} to formulate many-body theory in a wavepacket basis, rather than the standard extended Bloch basis used in long range ordered systems. To this end he introduced the orthonormal Wilson-Wannier basis \cite{Sullivan2010} with even and odd combinations of Wannier functions. Ossadnik \cite{Ossadnik2016} began by rewriting the many-body Hamiltonian in this wavepacket basis and proceeded to obtain a fermionic formulation of the SRO groundstate of the 2-leg Hubbard ladder at $\frac{1}{2}$-filling, in agreement with the standard bosonization analysis. He then went on to demonstrate that his wavepacket approach could be straightforwardly generalized to 2D, unlike bosonization techniques. However, he did not carry through explicit calculations for this case. In this paper we follow his lead and carry through such calculations in the simplest approximation scheme. The great advantage of the Ossadnik-WW formalism over a real space approach, is that it enables one to directly examine the behavior in $\mathbf{k}$-space near the Fermi energy. We can describe the opening of an energy gap in the 2-particle spectrum by U-processes in the antinodal region of $\mathbf{k}$-space. Analogously to the case of the Hubbard ladder, the presence of U-processes is the key to turning the superconducting gap at overdoping into an insulating pseudogap at underdoping.

A SRO model of fermions is characterized by short range correlations which decay exponentially at long distances. Ossadnik proposed a useful description of such behavior can be based on coarse graining the real space behavior leading to a lattice of supercells, whose length scales are determined by the range of the strong short range correlations. The long range behavior of the model is determined in the first place by the local excitations in a single supercell and then by inter-supercell interactions. Ossadnik noted that in this approach, the low energy spectrum of the individual supercell plays a crucial role in determining the form of the correlation at long distances. He illustrated this proposal by examining the contrasting behavior of two 1-dimensional $\frac{1}{2}$-filled Hubbard models, the single chain and the 2-leg ladder (2LL). In the former case a single supercell has a 2-fold degenerate groundstate and as a result inter-supercell interactions in the supercell lattice lead to a power law decay of long range AF correlations. In the case of a 2-leg Hubbard ladder, the supercell has an isolated groundstate with finite gaps to all excitations. As a result the weaker inter-supercell interactions act only to renormalize the finite magnitude of the single supercell excitation energies. He concluded that ``a surprising amount of information about the nature of the groundstate is present in the local interaction and excitation spectrum of the local problem in the WW basis'', i.e. the excitation spectrum of a single supercell. Ossadnik extended his real space coarse graining approach to 2-dimensions which leads to discretization of $\mathbf{k}$-space in the low energy region near the Fermi surface. Here we use this approach to describe the neighborhood of the 2-dimensional Fermi surface by a discrete set $\mathbf{k}$-space patches. We are motivated by the ARPES observations of a partially gapped Fermi surface in the pseudogap phase of underdoped cuprates with only SRO and no broken symmetry \cite{Yang2011}. Ossadnik based his analysis of the excitation spectrum of the 2LL by using the results of 1-loop RG calculations. This method however is limited to weak interactions. The excitation spectrum of the 2LL can be solved numerically using the DMRG method at all interaction strengths. This allows us to extend Ossadnik's analysis to more realistic values as will be discussed below.

\begin{figure*}
	[t] 
	\includegraphics[width=150mm]{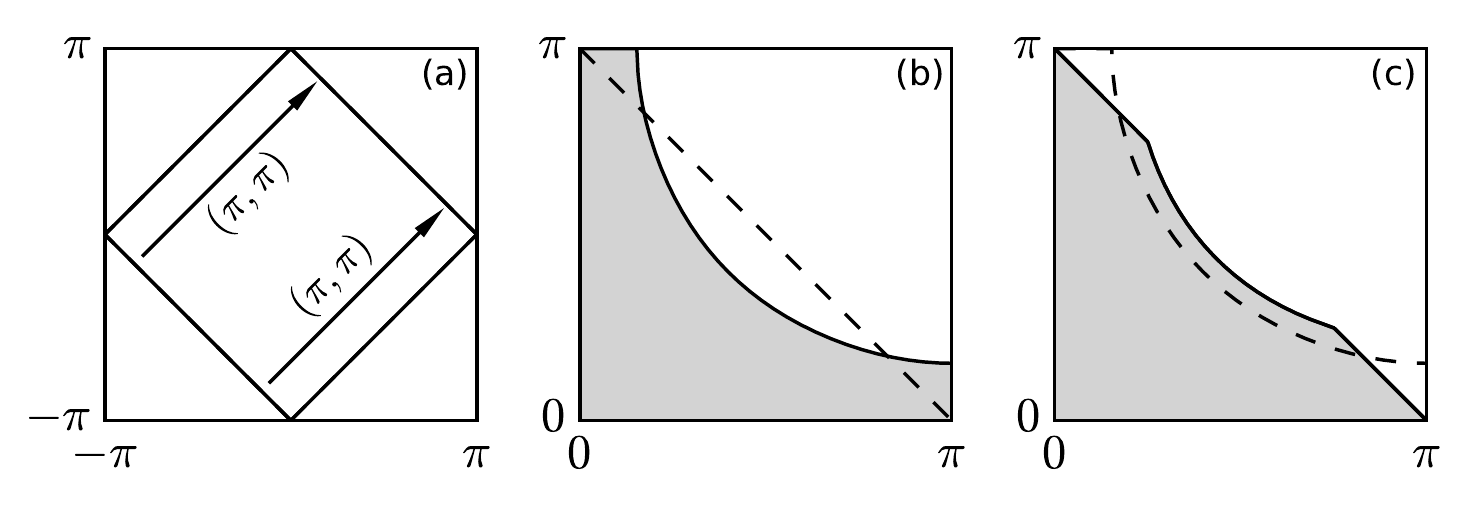} \caption{The YRZ reconstruction of the single particle distribution. The original Fermi surface (b) is deformed to the YRZ surface (c), by a transfer of occupied states from outside to inside of the US (a). This reconstruction conserves particle number and results in a small hole pocket near nodal (c).} 
\label{fig_YRZ} \end{figure*}

Our goal is to describe qualitatively the evolution in the physics, in particular the evolution of the 1-particle, 2-particle, and spin-flip excitation energies along the US. Alternative approaches have been put forward in a series of papers by Sachdev and coworkers \cite{Chowdhury2015}, Chubukov and collaborators \cite{Chubukov2015} and Pepin and coworkers \cite{Efetov2013} and many others. They start from the band structure Fermi surface and examine the role of U-scattering processes connecting the 8 ``hot points'', namely the 8 Fermi surface points lying on the US. They examine the possibility that as the hole density decreases below a critical value, coupled instabilities in the SDW and CDW channels can be stabilized in the Hubbard and spin fermion models. They interpret a number of X-ray investigations and Scanning Tunneling Microscope (STM) experiments in terms of such spatial modulations. We put forward a different interpretation here, which proposes that the key instability is driven not by a set of ``hot points'' on the original Fermi surface, but by a macroscopic reorganization of the $\mathbf{k}$-space ``Fermi surface'' to obtain ``hot lines'' lying on the US, see Fig.~\ref{fig_YRZ}. In a recent paper Liu and coworkers \cite{Liu2016} found that in this underdoped pseudogap state, the transition to superconducting long range order is preceded by a wide temperature region with strong phase fluctuations associated with a soft Leggett mode. Further they showed that this scenario explains the unusual Giant Phonon Anomalies reported by Le Tacon et al \cite{LeTacon2014} in high-resolution inelastic X-ray experiments on underdoped YBCO. In comparing these results on YBCO and the recent STM results on BSCCO \cite{Fujita2012} one must recognize the presence of strong static lattice disorder in the CuO$_{2}$ layers in BSCCO could account for the clear discrepancy between the two sets of experiments. This difference shows up very prominently when the strongly broadened NMR spectrum of a BSCCO \cite{Takigawa1994} sample is compared to the narrow well resolved lines in the NMR spectrum of underdoped YBa$_2$Cu$_4$O$_8$ \cite{Yasuoka1994}.

\section{The $U-J$ Hubbard Ladder} 

\begin{figure}
	[t] 
	\includegraphics[width=85mm]{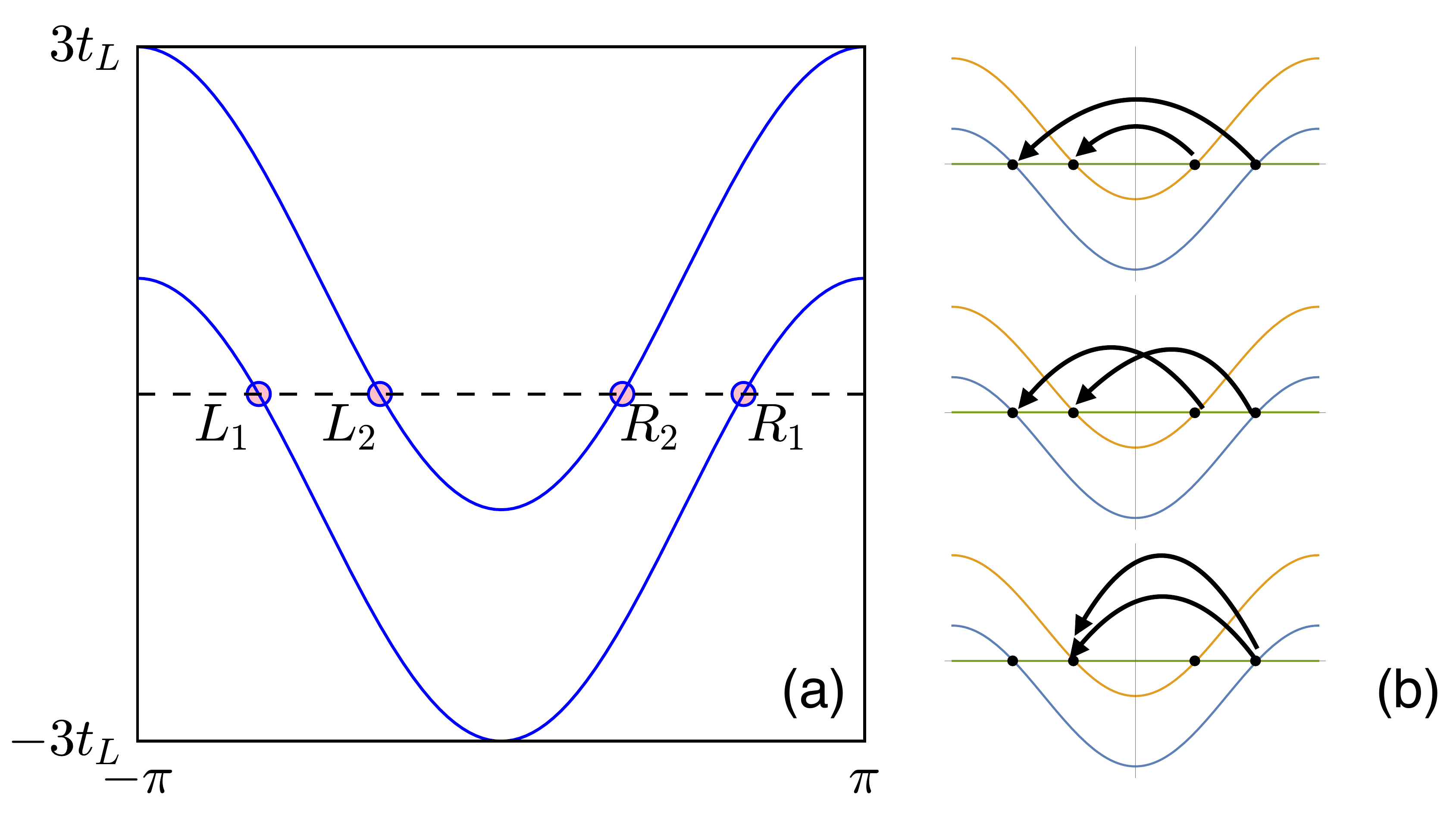} \caption{(a) The band structure of a 2-leg ladder at $\frac{1}{2}$-filling. The Fermi level crosses the bonding (1) and antibonding (2) rung bands leading to pairs of right (R) and left (L) moving Fermi points. Note at $\frac{1}{2}$-filling the Fermi points $R_1$ \& $L_2$ and $R_2$ \& $L_1$ are connected by $(\pi,\pm\pi)$ scatterings. (b) Umklapp scatterings at $\frac{1}{2}$-filling.} 
\label{fig_ladder} \end{figure}

We start by reviewing the U-scattering processes responsible for the D-Mott insulator at $\frac{1}{2}$-filling in a 2-leg Hubbard ladder. The band structure in Fig.~\ref{fig_ladder} has a pair of 1D bonding (1) and antibonding (2) bands with right and left moving Fermi wavevectors when we choose equal hopping matrix elements along legs and rungs, $R_1=(2\pi/3,0)$, $R_2=(\pi/3,\pi)$ and $L_1=(-2\pi/3,0)$, $L_2=(-\pi/3,\pi)$. The Fermi points at $R_1$ \& $L_2$ and also $R_2$ \& $L_1$, are separated by the wavevector $(\pi,\pm\pi)$ at $\frac{1}{2}$-filling, which as illustrated Fig.~\ref{fig_ladder}b, allows three new scattering processes with momentum change $(\pm2\pi,0)$ or $(\pm2\pi,\pm2\pi)$ at $\frac{1}{2}$-filling. The three extra interband scattering processes at $\frac{1}{2}$-filling change the nature of fixed point in 1-loop RG flow, to an insulator with both charge and spin gaps \cite{Balents1996}. In particular, a charge gap in the 2-particle Cooper pair channel is introduced, which converts superconducting correlations into strictly short range correlations. In the particle-hole channel the scattering of a particle-hole pair with opposite spins with an incoming wavevector of $R_1-L_2=(\pi,-\pi)$ is scattered to an outgoing particle-hole pair with wavevector $R_2-L_1=(\pi,\pi)$. This extra vertex is allowed only at the commensurate wavevector, which strengthens the SDW response at the commensurate $(\pi,\pi)$ wavevector. In the related particle-particle channel, this scattering process contributes to $d$-wave singlet pairing. Note however the commensurate order is pinned to the underlying lattice leading to a finite energy sliding mode in the particle-hole channel. The interference between the two SRO correlations can be constructive in the sense that it enhances the magnitude of the enegy gaps, but this comes at the cost of introducing an energy gap in the Cooper pair energy spectrum. The result is the novel SRO D-Mott insulator groundstate with a charge gap.

\begin{figure}
	[t] 
	\includegraphics[width=85mm]{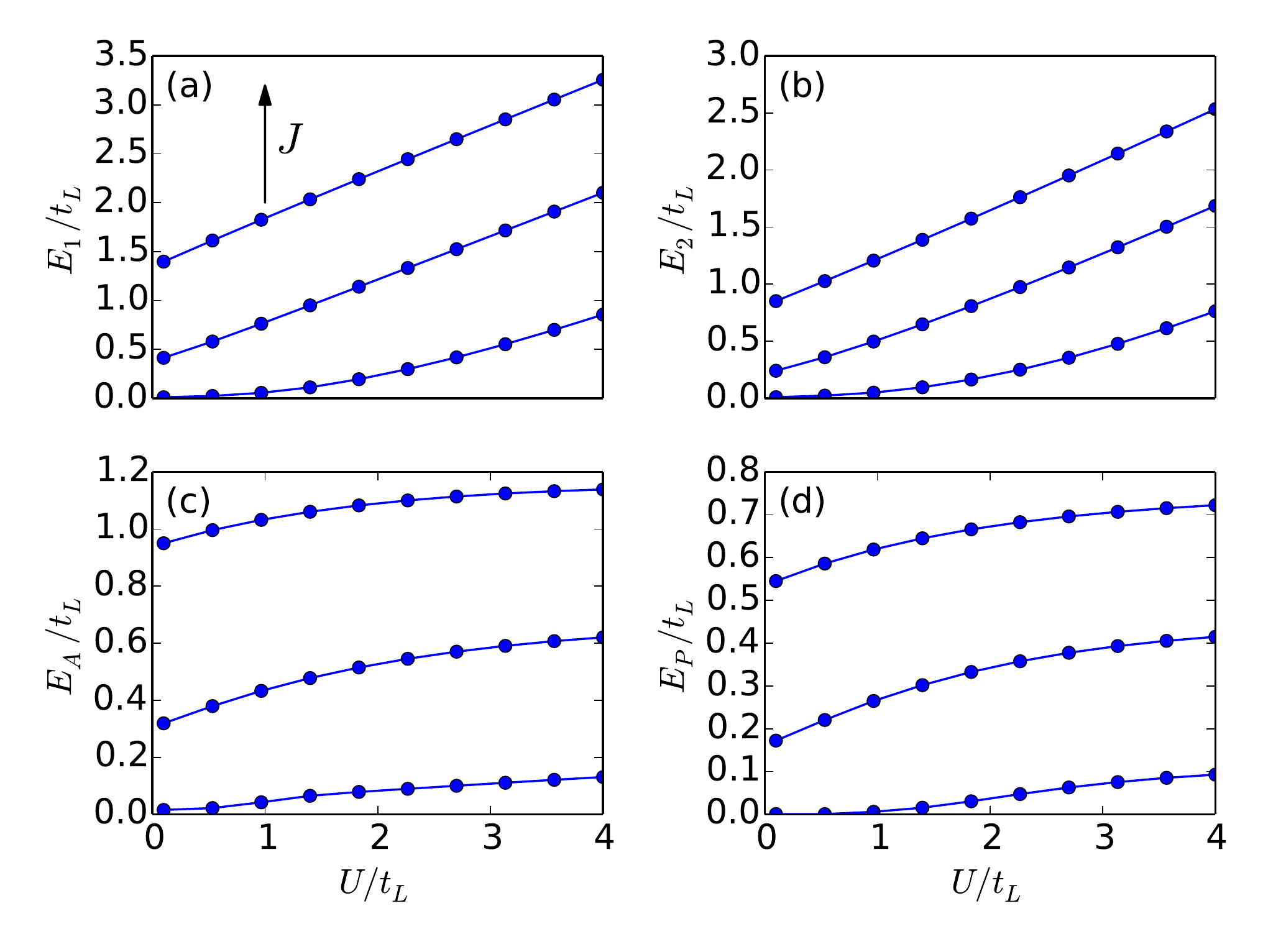} \caption{Excitation energies in the $U$-$J$ ladder at half filling. In each panel, the value of $J/t_L$ increases from bottom to top evenly from 0.1 to 2. All excitation energies increase smoothly as a function of $U$ and $J$, and the state is always the ``D-Mott'' insulator. However, the relative strength between SC and AF SROs could be tuned by $U$ and $J$. (Parameters: length $L=62$, bond dimension $D=700$.)} 
\label{fig_A2} \end{figure}

Before the discussion of the cuprate Fermi surface we examine the generalization of the D-Mott insulating state of the 2-leg Hubbard ladder to finite strength interactions. The 1-loop RG calculations are limited to weak coupling and the extension to stronger coupling by multi-loop RG is difficult. However numerical simulation of the 2-leg ladder by a real space DMRG method is possible \cite{Noack1994,Dolfi2015} and allows the calculation of the excitation energies at arbitrary strength interactions. In this section we summarize the numerical results. (The calculations are described in more detail in the Appendix.) 

We also examined the behavior when the interactions in the Hubbard model are extended by adding an antiferromagnetic spin-exchange interaction ($J$-term) to the original Hubbard onsite repulsion $U$. Note we are using the 2-leg Hubbard ladders to represent the low energy effective interactions after the high energy region of the 2D Hubbard model is integrated out. This process introduces uncertainty into the exact form of the resulting 1/2-filled 2-leg ladder in real space but as we shall see this does not lead to qualitative changes in the results. 
\begin{eqnarray}
	H &=& -t_L\sum_{\left<i,j\right>,\sigma} \left(c^\dagger_{i\sigma} c_{j\sigma} + c^\dagger_{j\sigma} c_{i\sigma} \right) \nonumber \\
	&&+U\sum_i n_{i\uparrow}n_{i\downarrow} + J \sum_{\left<i,j\right>} \mathbf{S}_i \cdot \mathbf{S}_j, 
\label{eq:model} \end{eqnarray}
where $\left<i,j\right>$ denotes nearest neighbors on the ladder in both leg and rung direction, $t_L$ is the hopping on the ladder (to distinguish from the hopping $t$ on 2D square lattice for later use), $\mathbf{S}_i=\frac{1}{2}\sum_{ab}c^\dagger_{ia}\vec{\sigma}_{ab}c_{ib}$ is the fermion spin. For this model, the strengths $U/t_L$ and $J/t_L$, can be varied to see the sensitivity of the results to variations in the low energy interactions along the umklapp surface. We use the DMRG code from the ALPS library \cite{Bauer2011,Dolfi2014} to simulate open ladders with length $L$ and width $W=2$, in various particle-number-conserving sectors with $N_\uparrow=\sum_i n_{i\uparrow}=L,L\pm1$ and $N_\downarrow=\sum_i n_{i\downarrow}=L,L\pm1$. Denoting the lowest eigenenergy in the sector $(N_\uparrow,N_\downarrow)$ by $E_{N_\uparrow,N_\downarrow}$, the 1-particle-, 2-particle- and spin-gaps, relative to the $\frac{1}{2}$-filled sector, are then defined by: 
\begin{gather}
	E_1 = \frac{1}{2}(E_{L+1,L}+E_{L-1,L})-E_{L,L}, \nonumber \\
	E_2 = \frac{1}{4}(E_{L+1,L+1}+E_{L-1,L-1})-\frac{1}{2}E_{L,L}, \nonumber \\
	E_A = E_{L+1,L-1}-E_{L,L}, \nonumber \\
	E_P=E_1-E_2. 
\end{gather}
Here we use $E_P$ to denote the energy gain when two particles are added together instead of separately, i.e. if $E_P>0$, then two particles have an effective attractive interaction, leading to Cooper pairing correlations. 

\begin{figure}
	[t] 
	\includegraphics[width=85mm]{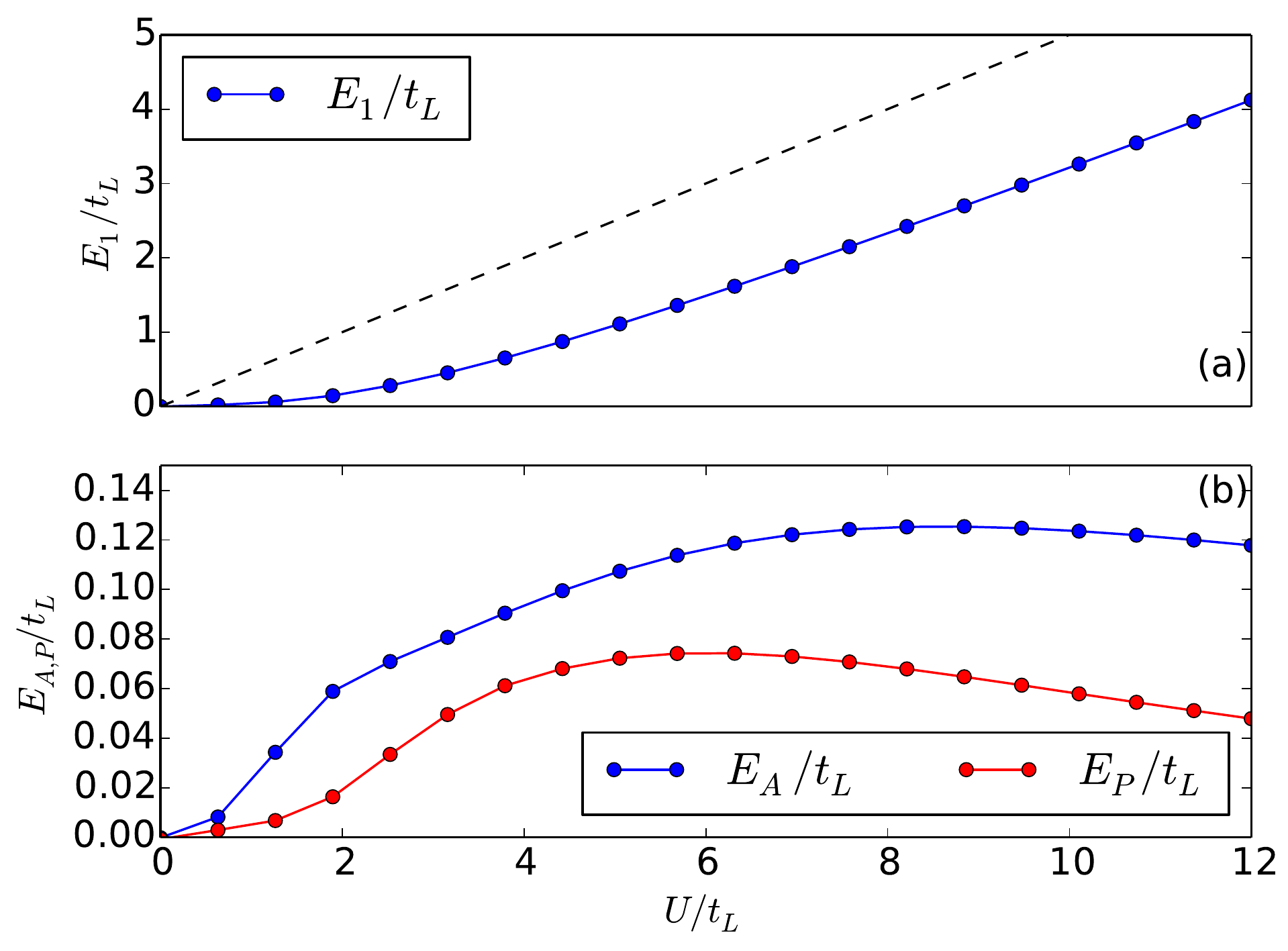} \caption{The excitation energies at $\frac{1}{2}$-filling, as a function of the interaction strength $U$. The charge gap grow monotonically with $U$, but the spin gap $E_A$ and pairing energy gain $E_P$ peak around a value of $U$ equal to the kinetic energy bandwidth $U \sim 6t_L$. (Parameters: length $L=62$, bond dimension $D=700$.)} 
\label{fig_Usweep} \end{figure}

\begin{figure}
	[t] 
	\includegraphics[width=85mm]{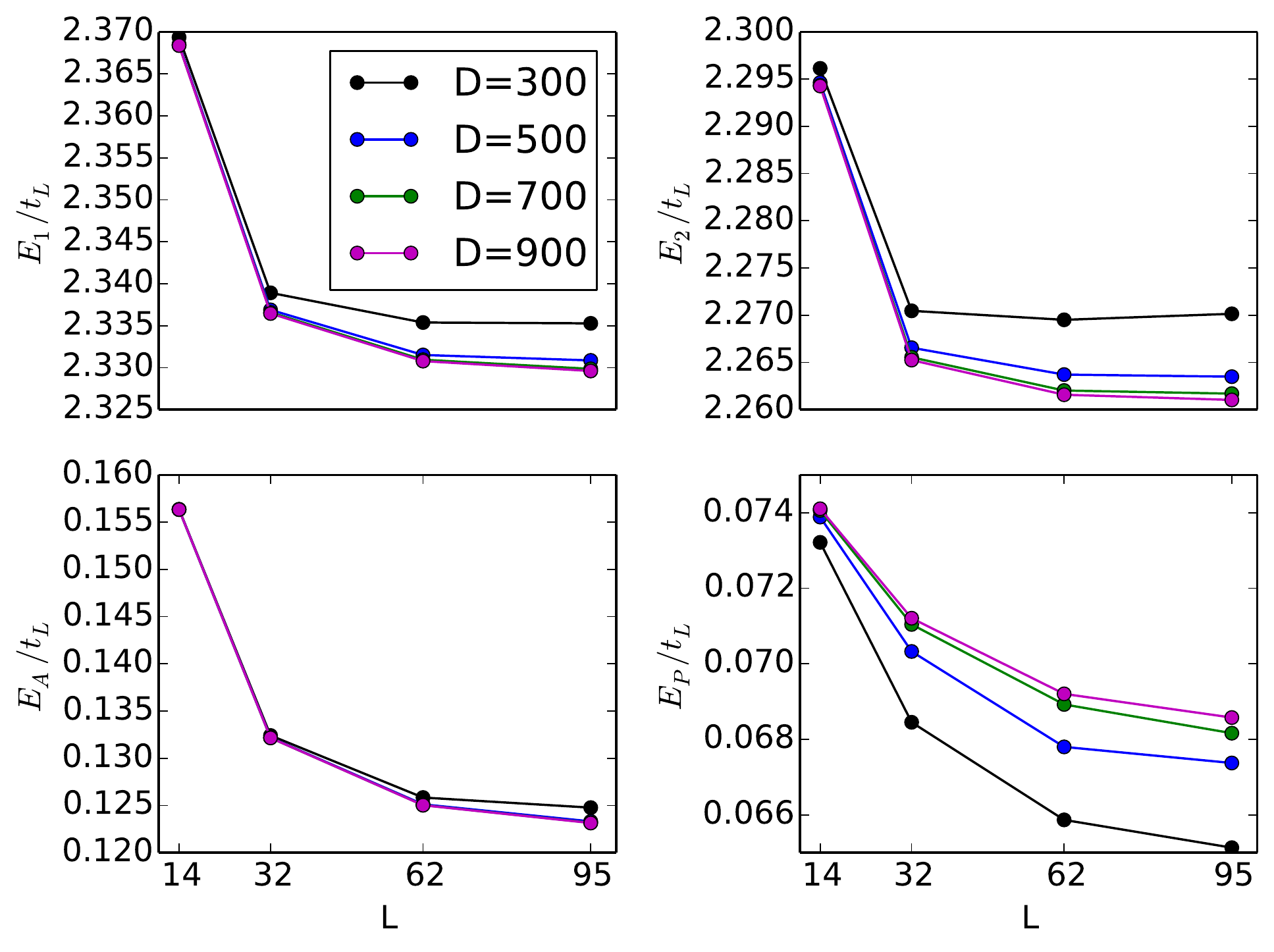} \caption{Finite-size and finite-bond-dimension effects on the excitation energies, for $U=8t_L$. The correlation length $\xi$ is defined by the length scale at which the excitation energies saturate. Further increases in the system size lead to only minor changes.} 
\label{fig_WW} \end{figure}

\begin{figure}
	[t] 
	\includegraphics[width=85mm]{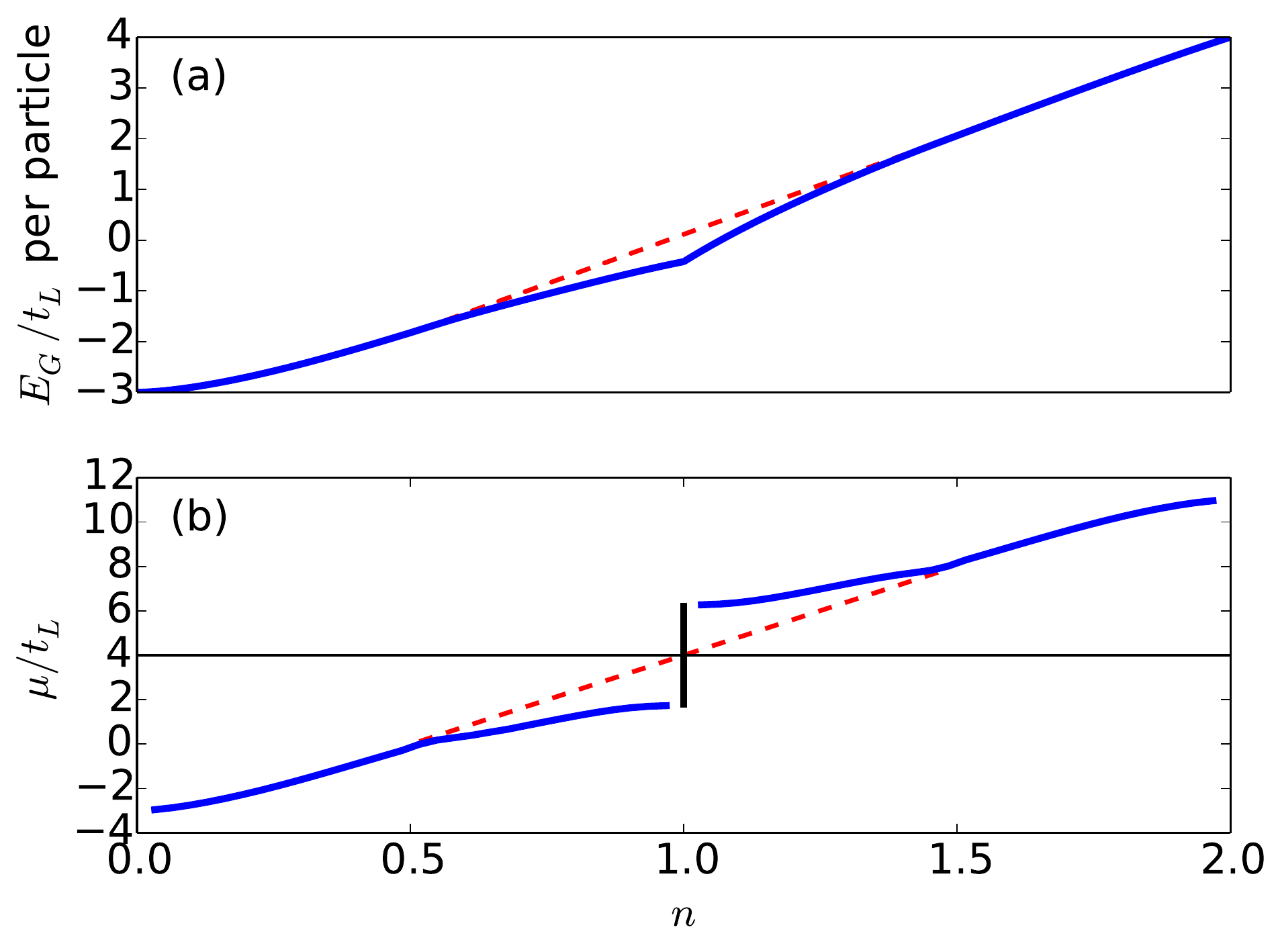} \caption{Groundstate energy per particle (a) and the chemical potential $ 
	\partial E_G / 
	\partial N$ (b), at $U=8t_L$ and $J=0$, as a function of filling, $n=N/2L$. The horizontal black line marks the particle-hole symmetric chemical potential $\mu=U/2=4t_L$, and the vertical black line the jump in $\mu$ due to the charge gap. (Parameters: length $L=62$, bond dimension $D=700$.)} 
\label{fig_Mott} \end{figure}
We begin by plotting the excitation energies at $\frac{1}{2}$-filling versus the parameters $U/t_L$ \& $J/t_L$ in Fig.~\ref{fig_A2}. We see a continuous rise of the excitation energies without any sign of a singularity. From this we conclude that the Mott physics in the 2-leg ladder at $\frac{1}{2}$-filling evolves smoothly from the weak coupling limit where it is caused by the extra umklapp processes at $\frac{1}{2}$-filling. This demonstrates that in 2D an insulating (or partially insulating) state can be realized without symmetry breaking if one can maximize elastic umklapp processes across parts of a Fermi surface.

The behavior of excitation energies, $E_1$, $E_2$ and $E_A$ with increasing interaction strength is also of interest (Fig.~\ref{fig_A2} and Fig.~\ref{fig_Usweep}). All three excitation energies are equal in the analytic solution in weak coupling \cite{Lin1998}. However, at finite coupling the single particle excitation is clearly larger than the energy per particle to add a Cooper pair. This behavior is robust and confirms the persistence of SRO in the pairing channel. The triplet $S=1$ excitation energy $E_A$ depends mostly on the value of $J$ and only weakly on $U$. The finite value indicates the persistence of AF $(\pi,\pi)$ SRO. 

The DMRG calculations also allow an explicit test of Ossadnik's central point, namely that SRO systems can be well represented by as a lattice of independent supercells in cases where the supercell groundstate is isolated with a finite energy gap for all excitations. To this end we show the finite-size effect on the excitation energies, for various ladder lengths. The results in Fig.~\ref{fig_WW}, do indeed show a clear saturation of the excitation energies at a finite length scale. This saturation length is smaller in the charge sector than the spin sector in agreement with the larger excitation energies in the former, so one can interpret the saturation length as the length scale of strong correlations. It also supports Ossadnik's proposal that inter-supercell interactions may be safely neglected. When we consider the 2D case below, we shall extend this approximation to neglect inter-supercell interactions also between supercells associated with correlations at different non-degenerate $\mathbf{k}$-points along the US.

Lastly we plot $E_G$ versus filling $n$, see Fig.~\ref{fig_Mott}. Here the solid blue lines are results from the DMRG, and the dashed red lines are interpolations as if there were no charge gap. As expected, the presence of elastic umklapp processes across the Fermi surface lowers the $E_G$ per particle. It follows that if a separate particle reservoir for particles exists, an extra single particle energy cost to transfer particles to the ladder, can be compensated by the energy gain due to the enhanced gap in $E_1$. Since the latter increases with increasing $U$ it follows that such a particle transfer will be stable beyond a critical value of $U$.

\section{The Pseudogap Phase in the Cuprates}

We use a single-band Hubbard model in 2D square lattice to describe the cuprates: 
\begin{align}
	H=-\sum_{ij\sigma}t_{ij}(c^\dagger_{i\sigma}c_{j\sigma}+c^\dagger_{j\sigma}c_{i\sigma})+U\sum_i n_{i\uparrow}n_{i\downarrow}, 
\end{align}
where $t_{ij}=t$ when $i,j$ are nearest neighbors (nn), and $t_{ij}=t'$ when $i,j$ are next-nearest neighbors (nnn). Later a third-nearest-neighbor hopping amplitude $t''$ (nnnn) is also included.

The original band structure Fermi surface of the cuprates in Fig.~\ref{fig_YRZ}b, only allows elastic U-scattering processes at the 8 so-called ``hot spots'' where the Fermi surface crosses the US. However since the pseudogap state is not present in the weak coupling limit, we should look also for interacting groundstates that derive adiabatically from noninteracting excited states. In the phenomenological YRZ ansatz to be discussed later \cite{Yang2006,Rice2012}, the original curved noninteracting Fermi surface is deformed to maximize the overlap with the US, as illustrated in Fig.~\ref{fig_YRZ}c. This excited band state is relevant if we postulate that the interacting groundstate is adiabatically connected to a noninteracting excited state with all occupied states inside the US. The boundary between occupied and unoccupied states extends along the US from the antinodal point until intersecting a Fermi arc centered on the nodal point of the US, as illustrated in Fig.~\ref{fig_YRZ}c. The commensurate AF state of Cr alloys is an example of a groundstate adiabatically connected to a Fermi surface modified to maximize U-scattering processes with wavevector $(\pi,\pi,\pi)$. This leads to greatly enhanced Neel temperature, energy gap and electron/atom range for commensurate itinerant antiferromagnetism in Cr alloys. The commensurate AF is adiabatically connected to an excited single-particle state, which leads to a doubling of the single-particle energy gap through the extra U-scattering processes that are allowed in the commensurate AF state \cite{Rice1970}. We note in passing that neutron scattering on cuprates with well ordered lattices, e.g. HgBa$_2$CuO$_{4+x}$, show only commensurate $(\pi,\pi)$ AF short range order in the pseudogap phase \cite{Chan2016}, an experimental confirmation that commensurate AF $(\pi,\pi)$ are playing an important role in stabilizing the pseudogap state. Note this change in the ``Fermi surface'' can be stabilized by a finite, but not necessarily strong, interaction strength at low energies as demonstrated by the case of Cr. In the cuprates as the doping decreases, the strength of the effective low energy repulsion increases, from zero in the Landau Fermi liquid at overdoping, to a large value in the Mott insulator with a large energy gap at zero doping. We shall return to the consequences for ARPES later.

Another feature of the pseudogap phase is the strong anisotropy in $\mathbf{k}$-space near the Fermi energy. The single-particle energy gap measured by ARPES is maximum near the antinodal points and decreases as $\mathbf{k}$ moves along the US towards nodal. In a hole doped sample, at a certain $\mathbf{k}$-point there is a crossover to a Fermi arc which lies inside the US and is centered at the nodal point. The superconducting gap that vanishes at $T_\mathrm{c}$ opens on this Fermi arc. 

Ossadnik compared this $\mathbf{k}$-space anisotropy near the Fermi energy to the behavior of multi-leg Hubbard ladders near $\frac{1}{2}$-filling. In this case the band structure is a set of 1D bands labelled by their transverse wavevector. These 1D bands form pairs with equal effective masses at the Fermi level, whose value depends on the rung wavevector \cite{Ledermann2000}. These band pairs have different density of states at the Fermi level and so make transitions to strong coupling at different critical scales in a RG treatment. This successive freeze out of the different band pairs as the energy scale is lowered suggests an approximation scheme, where we treat the freeze out of each band pair as independent as we lower the energy scale \cite{Ledermann2000}. As a result, in a lightly doped system at low temperature, we have a separation into $\frac{1}{2}$-filled band pairs with larger energy gaps, and lightly doped band pairs. This is similar in many ways to the variation with $\mathbf{k}$ along the US in the pseudogap phase in 2D. For a review of multi-ladder Hubbard models, see Le Hur and Rice \cite{leHur2009}.

\subsection{Trial wavefunction}

In the cuprates the 2D band structure energy varies along the US in the presence of nnn hopping, $t'<0$, with a minimum at antinodal and a maximum at nodal according to $-4t'\cos k_{x}\cos k_{y}$. Ossadnik showed the simplest way to treat the variation in $\mathbf{k}$-space along the US is to introduce a partition into independent $\mathbf{k}$-space patches, each characterized by a WW-wavevector and energy \cite{Ossadnik2016}. Each $\mathbf{k}$-patch belongs to a star of 8 degenerate $\mathbf{k}$-points on the US due to the square symmetry of the lattice. These points are connected by a set of scattering vertices including U-processes, which act to open extra energy gaps on the US, see Fig.~\ref{fig_YRZ}. Since the $\mathbf{k}$-stars have different energies and we are considering only short range order, we shall neglect scattering processes between different stars of 8 $\mathbf{k}$-patches. We represent the groundstate wavefunction as a product state of the groundstate of each 8 degenerate $\mathbf{k}$-point stars along the US. A further simplification occurs when we concentrate on the local approximation introduced by Ossadnik, namely we could evaluate the excitation energy spectrum of a single square supercell, and examine how this excitation energy spectrum varies with the position of the $\mathbf{k}$-patch on the US. The correlation lengths and hence the dimensions of each supercell vary with the position of $\mathbf{k}$ on the US. The shortest correlations on the fastest timescales will occur near the antinodal $\mathbf{k}$-points and the slower correlations grow in on longer length scales as $\mathbf{k}$ approaches nodal. 

\begin{figure}
	[t] 
	\includegraphics[width=65mm]{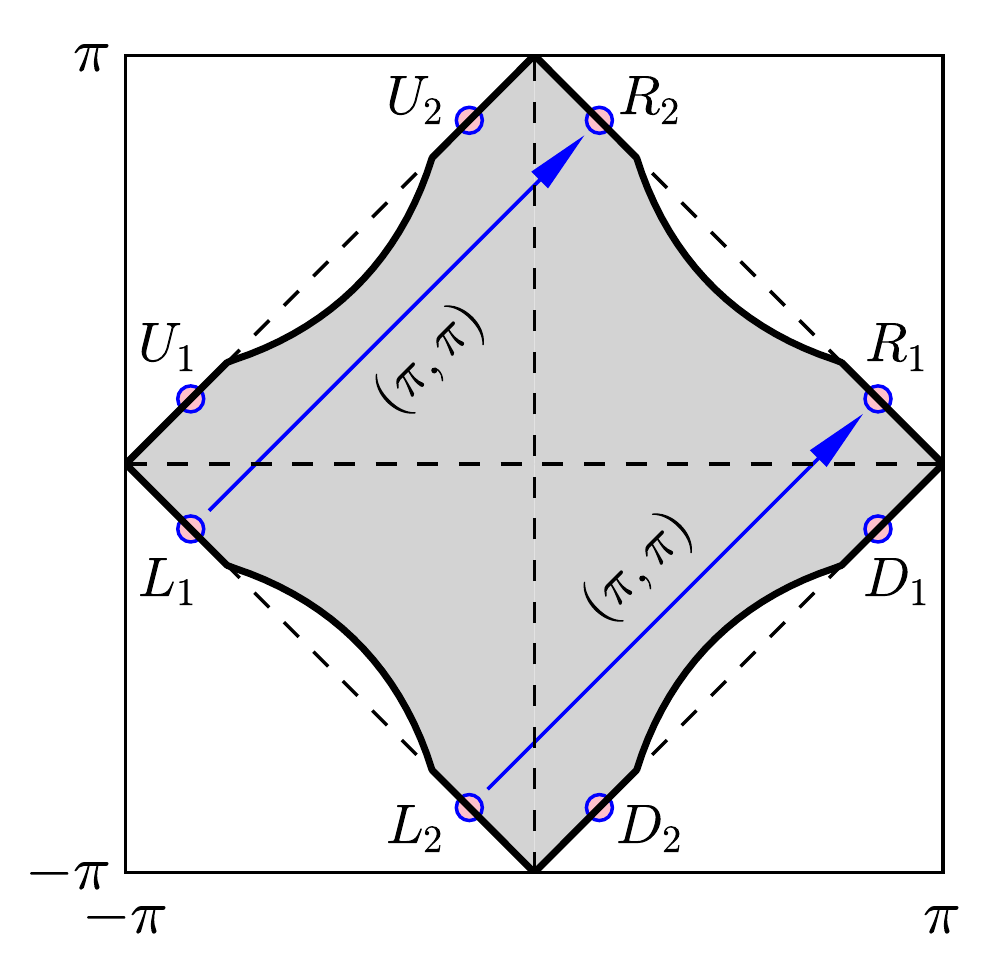} \caption{Mapping to the 8 degenerate ladder Fermi points. The 8 degenerate $\mathbf{k}$-points on the US (purple circles) are divided into into two subsets of 4 $\mathbf{k}$-points each, on the US. Each subset is connected by $(\pi,\pi)$ intra-subband scattering, similar to the case of 2-leg ladders in Fig.~\ref{fig_ladder}. The gray shaded region is occupied by the YRZ construction. The blue arrows mark the important intra-subband $(\pi,\pi)$ scatterings.} 
\label{fig_map} \end{figure}

A typical star of 8 $\mathbf{k}$-points is represented in Fig.~\ref{fig_map}. For a $\mathbf{k}$-space analysis we need to parameterize the dominant scattering processes at low energies in a 2D Hubbard model when the Fermi surface is near the US. Under these conditions there are competing and mutually interacting collective fluctuations in both spin-density-wave and $d$-wave-pairing channels. The singular-mode FRG (SM-FRG) \cite{Husemann2009,Xiang2012} represents the effective interactions as a coupled flow of wavevector resolved terms in the $d$-wave pairing ($d$SC) and particle-hole AF $(\pi,\pi)$ channels. While there are explicit calculations only to 1-loop order we shall apply this form to moderately strong interactions relevant in the pseudogap phase. This form of the interaction is similar to that proposed for strong coupling in the ``spin-Fermion'' model \cite{Abanov2003,Wang2013}. 

Motivated by the Ossadnik-WW theory, we separate the conducting and insulating parts in $\mathbf{k}$-space, and propose the pseudogap wavefunction to be a product of the BCS state on the nodal arc, and an insulating state on the umklapp surface near antinodal. The insulating state is in turn a product state of different sets of 8 Fermi points (patches). The wavefunction reads: 
\begin{align}
	\Ket{\Psi}_\mathrm{BZ} = \Ket{\mathrm{BCS}}_\mathrm{arc} \otimes \Ket{\mathrm{PG}}, \, \Ket{\mathrm{PG}} = \bigotimes_{\mathbf{k}_\mathrm{U}}\Ket{\mathrm{2LL}}_{S(\mathbf{k}_\mathrm{U})}, 
\label{eq:psi} \end{align}
which is in a product form. Here BZ denotes the whole Brillouin zone, $S(\mathbf{k}_\mathrm{U})$ denotes one set of 8 Fermi points connected from an umklapp point $\mathbf{k}_\mathrm{U}$, e.g. in Fig.~\ref{fig_map}, $S(R_1)=\{L_1,L_2,R_1,R_2,U_1,U_2,D_1,D_2\}$. In passing we note that as far as we are aware, so far an approximate wave function, which includes US has not been found.

Note the product form in $\mathbf{k}$-space for the 2D groundstate has certain similarities to the product form of a BCS wave function. However in that case, all $\mathbf{k}$-patches are degenerate and the presence of long range order creates a coherent field on each $\mathbf{k}$-patch due to inter-$\mathbf{k}$-patch interactions. In the pseudogap phase there is a continuous crossover to SRO as the temperature is lowered making it similar to case of wider Hubbard ladders at 1/2- filling, where successive band pairs cross into the strong coupling state as the energy scale is lowered. 

\subsection{Mapping to 2LLs}

Looking at the 8-$\mathbf{k}$-point star illustrated in Fig.~\ref{fig_map}, we see that it can be decomposed of 2 subsets $(R,L)$ \& $(U,D)$ with velocities along the $(1,1)$ \& $(1,-1)$ directions respectively. Each subset consists of 4 $\mathbf{k}$-points connected by umklapp scattering processes involving $(\pi,\pi)$ momentum transfers, similar to the case of 2LLs at $\frac{1}{2}$-filling. While there are umklapp scattering processes between $\mathbf{k}$-points in the two different subsets, these U-processes do not involve $(\pi,\pi)$ momentum transfers. This suggests a further simplification, where we approximate the groundstate of the 8-$\mathbf{k}$-point star as the product of a pair of independent 4-$\mathbf{k}$-point subsets with velocities along the $(1,1) $ \& $ (1,-1)$ directions. This product approximation greatly simplifies the calculations as each 4-$\mathbf{k}$-point subset can then be mapped onto a 2LL.

This connection is clear when we label the 4 degenerate $\mathbf{k}$-points of a single subset on the US as right ($R$) and left ($L$) movers in Fig.~\ref{fig_map}, and add a second label 1 \& 2 to denote the $\pm\mathbf{k}$ pairs. The similarity of this 4-$\mathbf{k}$-point subset to the 4 Fermi points of the 2LL at $\frac{1}{2}$-filling in Fig.~\ref{fig_ladder}, is obvious. The wavevector separations between $R_1$ \& $L_2$, and also $R_2$ \& $L_1$, are $(\pi,\pi)$, similar to the 2-leg ladder at $\frac{1}{2}$-filling so that a scattering between the Cooper pairs $(R_1$,$L_1)$ \& $(R_2,L_2)$ involves momentum transfers of $(\pi,\pi)$, just as in the 2LL. The spin response is peaked at the commensurate wavevector $(\pi,\pi)$, enhancing the strength of scattering processes through this wave vector. As discussed earlier the analytic solution in $\mathbf{k}$-space for ladders is limited to weak coupling, but a numerical solution using DMRG can be carried out in a real space formulation for arbitrary coupling. This requires a real space, rather than $\mathbf{k}$-space form for the interaction e.g. represented as $U,J$ in Eq.~(\ref{eq:model}). The relevant values of $U,J$ are not easy to estimate but as shown earlier the excitation energies $E_1,E_2$ and $E_A$ at $\frac{1}{2}$-filling evolve smoothly with $U,J$. Thus our approximate treatment will not be quantitive, only qualitative. 

\begin{figure*}
	[t] 
	\includegraphics[width=180mm]{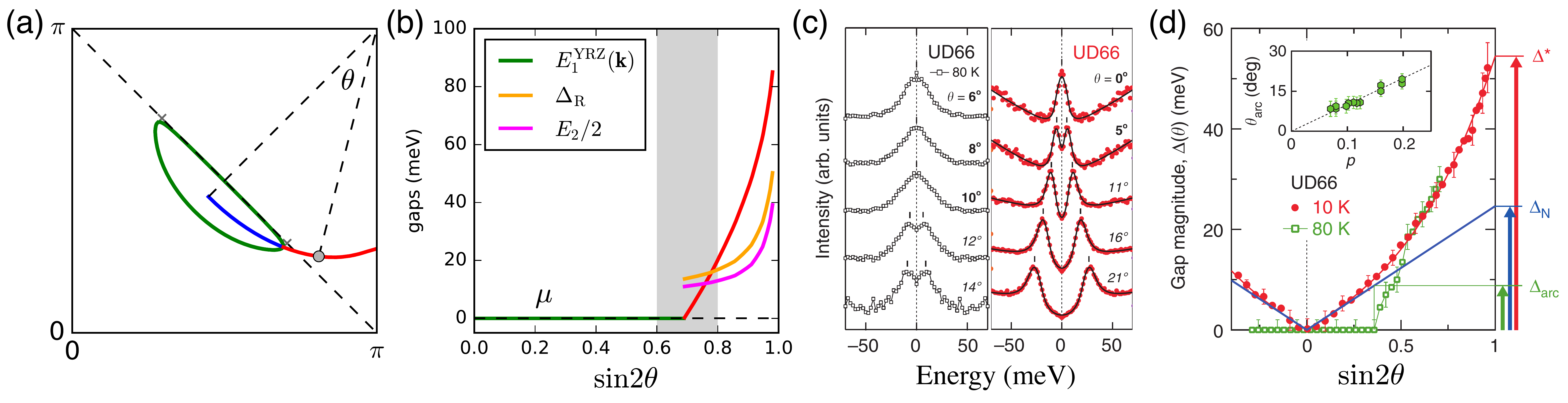} \caption{ (a) Normal state energy contours, $E_1(\mathbf{k})$ in YRZ, form a cone (blue) above the nodal pocket Fermi surface (green). The contour (red) beyond the pocket end denotes the minimum energy to extract an electron. Note, the $E_1(\mathbf{k})$ contour deviates from the US due to the non constant electron energy on the US. (b) Energy values along these contours. Outside the pocket (larger $\theta$) the pseudogap hole energy, $E_1$ (red) increases to the Brillouin zone boundary, see Yang et al \cite{Yang2011}. The YRZ gap parameter $\Delta_\mathrm{R}$ (estimated from DMRG ladder calculations) is shown in orange. In the pseudogap region, the energy per hole $E_2$ (magenta), to extract a Cooper pair of holes (estimated from DMRG ladder calculations) lies below $E_1(\mathbf{k})$ due to SRO pairing correlations. The transition region between the end of the Fermi pocket and the onset of the pseudogap is shaded gray. In the normal state at $T > T_\mathrm{c}$ it is characterized by strong pairing fluctuations. (Parameters used in the DMRG calculations: $U=0.4t$, $J=0.1t$, $L=62$, $D=900$. Parameters for YRZ: $t_0=0.3t$, $t_1=-0.06t$, $t_2=0.08t$, $\mu_\mathrm{P}=-0.089t$. See the appendix for more details.) (c) The energy symmetrized ARPES results for $E_1(\mathbf{k})$ reported by Anzai et al \cite{Anzai2013} at both $T = 10$K (red) in the superconducting state, and $T = 80$K (black) in the normal state on underdoped BSCCO ($T_\mathrm{c}=66$K). Note the energy splitting at $T=10$K changes to a small splitting between two broad peaks at $T = 80$K, due to pairing fluctuations at $T > T_\mathrm{c}$. (d) A comparison of the values of the single hole extraction energies, $E_1(\mathbf{k})$ at low $T$ (10K) in the superconducting state and high $T$ (80K) in the normal phase. } 
\label{fig_ARPES} \end{figure*}

\subsection{Relation to the YRZ ansatz for the single particle propagator and ARPES experiments}

We begin by examining the $E_1(\mathbf{k})$, which is measured in ARPES experiments. Numerous experimental investigations of the electronic properties in the pseudogap state have found a strong dichotomy between the $\mathbf{k}$-space regions near antinodal and nodal. Only SRO correlations accompanied by insulating behavior are observed near antinodal, which persist up to high temperatures. This behavior is a clear contrast with the large superconducting gap at $T<T _\mathrm{c}$, observed near antinodal at overdoping. We have argued above that this difference is similar to that observed in the SRO insulating state driven by umklapp processes observed at $\frac{1}{2}$-filling in a 2-leg Hubbard ladder. Here this behavior was derived when the starting band structure Fermi surface is modified to maximize overlap with the US. In this case the dominant interactions are those within the degenerate 4-$\mathbf{k}$-point subsets with local velocities along the $(1,1)$ \& $(1,-1)$ directions. Further we argued the U-processes connecting the two 4-$\mathbf{k}$-point subsets are weaker, allowing us to apply results from the SRO insulator driven by umklapp processes observed at $\frac{1}{2}$-filling in a 2-leg Hubbard ladder.

Some years ago, Yang, Rice and Zhang \cite{Yang2006} put forward a phenomenological form for the single particle propagator whose poles determine $E_1(\mathbf{k})$, stimulated by the prominence of U-scattering processes in the earlier FRG results of Honerkamp et al \cite{Honerkamp2001}: 
\begin{align}
	G^\mathrm{YRZ}(\mathbf{k},\omega) = &\frac{g_t}{\omega-\xi(\mathbf{k}) - \frac{\Delta_\mathrm{R}^2(\mathbf{k})} {\omega+\xi_0(\mathbf{k})}} + G_\mathrm{inc.},\nonumber\\
	\xi_0(\mathbf{k}) =& -2t(\cos k_x+ \cos k_y), \nonumber\\
	\xi(\mathbf{k}) =& \xi_0(\mathbf{k})-4t'\cos k_x \cos k_y\nonumber\\
	&-2t''(\cos2k_x+\cos2k_y) - \mu_\mathrm{P},\nonumber\\
	\Delta_R(\mathbf{k}) =& \Delta_0(\cos k_x-\cos k_y). 
\end{align}
The form of the self energy is a generalization of the self energy of a single $\frac{1}{2}$-filled 2LL at $T=0$K, obtained by Konik and Ludwig \cite{Konik2001}. The presence of both charge and spin gaps leads to a well defined pole at each of the four $k_F$ points and a divergence of the self energy $\Sigma$ at $\omega \rightarrow 0$ at these points. The YRZ self energy has a similar divergence in the self energy along the US. This form can be tested by ARPES experiments, which measure the minimum energy to create a hole in the occupied states, $E_1(\mathbf{k})$. Figure~\ref{fig_ARPES}a shows the energy contours of $E_1(\mathbf{k})$ in the absence of superconductivity. In the nodal region, $E_1(\mathbf{k})$ forms an anisotropic Fermi pocket shown in green. Above the chemical potential this pocket closes with a local maximum along the blue line. Outside the nodal pocket the minimum hole energy, $E_1(\mathbf{k})$, lies on the red curve. Note this locus deviates from the US due to the nnn hopping term which causes the band structure energy to vary on the US. This deviation is present in ARPES experiments \cite{Yang2011}. Figure~\ref{fig_ARPES}b illustrates the energies of create a hole along the red contour connecting the Fermi pocket to the BZ boundary. Also shown is the gap $\Delta(\mathbf{k})$ and its evolution along the US postulated in the YRZ propagator. This determines $E_1(\mathbf{k})$ and $-\frac{1}{\pi}\mathrm{Im}{G^\mathrm{YRZ}(\mathbf{k},\omega)}$ and the ARPES line shape. In the original YRZ paper a simple $d$-wave form for $\Delta(\mathbf{k})$ was assumed. 

The product form Eq.~(\ref{eq:psi}) for the groundstate wavefunction gives a microscopic derivation of $E_1(\mathbf{k})$ for the 2D system. In this case there is a continuous distribution of $\mathbf{k}$-points along the US, which we can obtain by interpolating between the $\mathbf{k}$-point patches. Note that the dimensionless parameters $U/t_L,J/t_L$ in the mapping to 2LL in Eq.~(\ref{eq:model}) involve the DOS at the US, which decreases along the US from antinodal towards nodal. As a result the effective ladder hopping, $t_L$, decreases and this in turn causes a decrease in $E_1(\mathbf{k})$ as $\mathbf{k}$ moves from antinodal towards nodal, a behavior that is also found in the YRZ ansatz. The mapping to 2LL goes beyond the YRZ ansatz and determines the value of $E_2(\mathbf{k})$, the energy per hole to extract a Cooper hole pair. It gives a finite value for $E_2(\mathbf{k})$ in the antinodal region below $E_1(\mathbf{k})$ which causes the insulating character of the antinodal pseudogap. Note however that $E_2(\mathbf{k})$ < $E_1(\mathbf{k})$ due to the introduction of SRO of Cooper pairs in the pseudogap state. This prediction however can not be directly tested by experiment.

\begin{figure}
	[t] 
	\includegraphics[width=85mm]{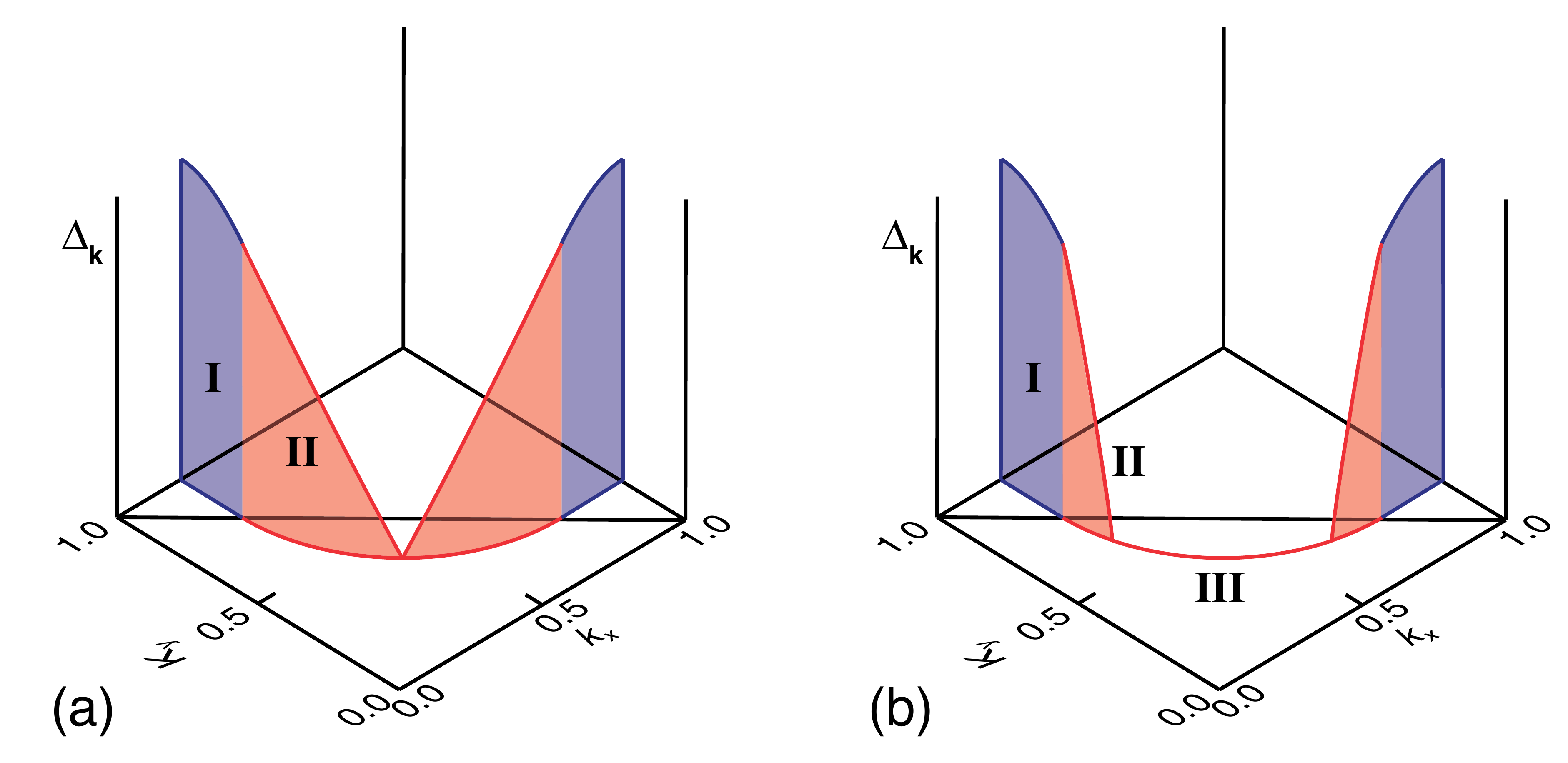} \caption{Scanning Tunneling Microscopy results for $E_1(\mathbf{k})$ by Fujita et al (Ref.~\cite{Fujita2012}). (a) I, pseudogap region; II, superconducting region. (b) I, pseudogap region; II, transition region; III, Fermi arc.} 
\label{fig_cite_JCD} \end{figure}

In the panels Fig.~\ref{fig_ARPES}c and Fig.~\ref{fig_ARPES}d, we reproduce a recent set of ARPES experiments by Anzai et al \cite{Anzai2013}, showing the temperature dependence of hole spectra, $E_1(\mathbf{k})$, in an underdoped BSCCO sample with a superconducting $T_\mathrm{c} = 66$K. The increase in the magnitude of $E_1(\mathbf{k})$ as $\mathbf{k}$ moves towards antinodal is clear in ARPES results. Note the absence of a change in $E_1(\mathbf{k})$ near antinodal, between the superconducting state at $T = 10$K and the normal state at 80K, is consistent with the irrelevance of superconducting LRO to the gap in the antinodal region. This behavior is to be expected since the pseudogap persists to $T^*$, substantially higher than $T_\mathrm{c}$. In this region, the minimum energy to extract a hole is displaced away from the antinodal $\mathbf{k}$-point in agreement with the result from the YRZ form for the single particle propagator. The locus of the minimum in $E_1(\mathbf{k})$ is plotted in Fig.~\ref{fig_ARPES}d agrees with that found by ARPES \cite{Yang2011}. The U-processes not only contribute to $E_1(\mathbf{k})$ but open a gap also in the pair spectrum $E_2(\mathbf{k})$ leading to an insulating antinodal pseudogap.

The superconducting order at low $T$ opens a $E_1(\mathbf{k})$ gap on the anisotropic metallic nodal pockets. The YRZ ansatz applies only to the normal phase and an additional residual attractive $d$-wave pairing interaction has to be added to explain the superconducting transition at $T_\mathrm{c}$. The metallic nodal pockets are displaced away from the US in Fig.~\ref{fig_YRZ}c, this suppresses U-processes and allows the onset of superconductivity in the metallic pockets. An overall reduction of $T_\mathrm{c}$ associated with the onset of the pseudogap follows from the decrease in the length of the metallic Fermi surface. The transition to superconducting order at $T_\mathrm{c}$ opens up a single particle gap $E_1(\mathbf{k})$ on the Fermi surfaces of the nodal pockets, as confirmed by the ARPES experiments, e.g. in Fig.~\ref{fig_ARPES}d. As usual, $E_2(\mathbf{k})$ is zero in a superconductor.

The dichotomy between the electronic spectra in the antinodal and nodal regions of $\mathbf{k}$-space is well reproduced in the 2LL approximation through the presence and absence of strong U-scattering processes, respectively. There remains however the transition region between these regions, i.e. the insulating pseudogap on the US and the end of the Fermi arcs. At low temperatures in the superconducting state, the ARPES data in Fig.~\ref{fig_ARPES}d show a distinct kink at the transition between the two $\mathbf{k}$-regions. This can be rationalized by the presence of extra U-scattering at the transition from the metallic Fermi surface to the US.

At $T = 80$K in the normal state, the ARPES results by Anzai et al \cite{Anzai2013}, in Fig.~\ref{fig_ARPES}d show a transition region in $E_1(\mathbf{k})$ joining a normal Fermi surface near nodal to the temperature independent antinodal region. However a closer inspection of the individual ARPES spectra in Fig.~\ref{fig_ARPES}c show a substantial broadening of the peaks in the ARPES spectra at $T = 80$K over those at $T=10$K, indicating strong superconducting fluctuations in the pairing state

Another way to measure $E_1(\mathbf{k})$ is through Spectroscopic Imaging Scanning Tunneling Microscopy. Recent results by the Davis group \cite{Fujita2012} for $E_1(\mathbf{k})$ on underdoped BSCCO in the pseudogap phase, for both the long range ordered $d$SC and normal ($T =1.5T_\mathrm{c}$) states, are summarized schematically in Fig.~\ref{fig_cite_JCD}. In the antinodal region I, the gapped excitations are associated with the finite pseudogap. The region marked II in the $d$SC superconducting phase at $T < T_\mathrm{c}$ is interpreted as arising from Bogoliubov quasiparticles associated with delocalized Cooper pairs, similar to our previous discussion. At $T >T_\mathrm{c}$, region III is the normal metallic Fermi arc and region II is interpreted in terms of $d$SC fluctuations, which appear at the transition region between the nodal and antinodal regions.

There have been a number of other independent measurements reporting superconducting fluctuations existing over a large temperature region at underdoped hole densities. Dubroka et al \cite{Dubroka2011} found evidence in the $c$-axis infrared spectra of a transverse Josephson plasmon in underdoped YBCO extending up to $3T_\mathrm{c}$. Recently Koren and Lee \cite{Koren2016} found anomalous tunneling spectra in epitaxial $c$-axis junction which they interpreted as evidence for preformed pairs at $T > T_\mathrm{c}$. Le Tacon et al \cite{LeTacon2014} observed Giant Phonon Anomalies in Inelastic X-ray Spectroscopy, whose form was greatly modified at $T_\mathrm{c}$. Liu et al \cite{Liu2016} found these unusual phonon anomalies could be explained by the existence of superconducting fluctuations associated with Leggett mode fluctuations in the normal state above $T_\mathrm{c}$. They derived an unusual Leggett mode resulting from strong internal pairing interactions inside the the Fermi arc pairs in the $(1,1)$ \& $(1,-1)$ directions and weak pairing between these two sets of arc pairs, similar to that proposed above. Their analysis leads to very strong fluctuations as $T_\mathrm{c}$ is approached from above, which couple to phonons with wave vectors connecting the arc ends. These are the wave vectors where Le Tacon et al \cite{LeTacon2014} found the Giant Phonon Anomalies. 

The third excitation energy is $E_A$ -- the spin triplet excitation with energy minimum at $(\pi,\pi)$. We won't discuss this in detail here. The ladder results show that this is mostly dependent on the strength of the $J$-term. Neutron scattering results on the clean underdoped single layer cuprate, Hg1201, show a clear minimum in $E_A$ at $(\pi,\pi)$ but without the lower energy legs reported in the ``hourglass'' form in several 214 cuprates \cite{Chan2016}. These low energy legs presumably come from particle-hole transitions between opposite arcs. Their absence in Hg1201 may be due to the much weaker AF correlations for these transitions which do not connect $\mathbf{k}$ on the US compared to those involving the US sections with strong AF 2LL correlations.

\section{Discussion}

In this paper we have proposed a generalization of the well studied D-Mott insulator groundstate of the $\frac{1}{2}$-filled 2-leg Hubbard ladder to 2D. The truncation of the Fermi surface in the ladder groundstate by correlations that are strictly short range, even at weak coupling, leads us to look into the possibility of a generalization to 2D. Such a state is a promising candidate to describe the instability of the overdoped full Fermi surface metal to the pseudogap state, as the hole density is reduced. Ossadnik's recent wavepacket formulation of many body theory offers a way to investigate modifications of electronic properties through strictly SRO correlations. The key feature of the D-Mott insulator in 1D at weak coupling is the presence of elastic umklapp scattering processes spanning the Fermi surface. Generally in higher dimensions the surface spanned by elastic umklapp processes has little overlap with the Fermi surface. So a precursor Mott state is only possible at strong coupling. However the 1D example of the $\frac{1}{2}$-filled 2-leg ladder suggests that there could be special Fermi surfaces that are susceptible to truncation by umklapp processes even at moderate coupling. The relatively small deviations between the 2D Fermi surface of the CuO$_2$ layers in cuprates and the surface that is spanned by umklapp processes in a 2D square lattice near $\frac{1}{2}$-filling led us to investigate this possibility further. In particular we found a nearby excited band structure state with a special ``Fermi'' surface separating occupied and empty states. This situation is reminiscent of the case of Cr alloys, where a small distortion of the original Fermi surface strongly enhances commensurate nesting thereby enhancing the stability of the commensurate SDW phase. In the cuprates it is the commensurate reduction of the hole density from its value of $1 + x$ in the band structure state at overdoping to realize a precursor Mott state with hole doping $x$ at a value $x\sim0.19$.

The model we considered here is closely related to the earlier YRZ model \cite{Yang2006}. Their ansatz postulated a form for the self energy similar to that in the $\frac{1}{2}$-filled 2-leg Hubbard ladder \cite{Konik2001}. Both focus on insulating energy gaps opening on the US rather than on the overdoped Fermi surface. The earlier pairing ansatz \cite{Zhang1988,Yang2006} was inspired by a mean field factorization of the strong coupling $t$-$J$ Hamiltonian and the crossover to the full Fermi surface at overdoping appeared when the finite solution to the gap equation vanished in a continuous transition between underdoping and overdoping. In the present approach the transition is due to an energy crossing between two different groundstates with energy gaps on the US and on the band structure Fermi surface. In the simplest description this leads to a first order transition although this may appear in real materials as continuous due to local fluctuations in the hole density caused by spatial disorder of the acceptors. One difference between the two proposals is the appearance of near antinodal electron pockets in YRZ as the transition is approached from underdoped. As far as we are aware, the presence of these near antinodal electron pockets have not been seen directly in experiments.

Recently a detailed study of the low-field Hall effect through the transition has been reported by Badoux et al \cite{Badoux2016}. They find an initial sharp drop in the carrier density from the full Fermi surface value at overdoping with longer tail on the low hole density side. This sharp drop could be considered evidence for a discontinuity in carrier density. Note however the crossover agrees well with Storey's calculation \cite{Storey2016} using the YRZ ansatz. As a result a firm conclusion between a continuous transition a la YRZ and a discontinuity possibly broadened by spatial disorder, is not possible at present.

Finally we comment on the nature of lattice fluctuations in the pseudogap state at temperatures $T > T_\mathrm{c}$. Many authors have looked to explain these as due to the presence of Charge Density Waves associated in some way with the SDW fluctuations and also as competing with superconductivity in this temperature range \cite{Hayward2014,Efetov2013,Bulut2013,Melikyan2014,Fradkin2014, Tsvelik2014,Chowdhury2015,Wang2015,Lee2014}. CDW and phonon fluctuations do not enter in the theoretical model presented here. A recent paper by Liu et al. \cite{Liu2016} gave an alternative explanation for the experiments on this topic, based on the presence of an extended range of superconducting fluctuations above superconducting $T_\mathrm{c}$ in the pseudogap state. Using arguments similar to those that are used here to justify the separation between the $(1,1)$ \& $(1,-1)$ movers, they argued for strong phase fluctuations between the pairing amplitudes on the arcs in these directions are accompanied by overdamped Leggett modes. They found that these Leggett modes can couple to lattice phonons and lead to unusual Giant Phonon Anomalies in agreement with those observed by Le Tacon et al \cite{LeTacon2014} on underdoped YBCO with a $T_\mathrm{c} \sim 60$K. Their special feature is a strong increase in the broadening of phonons at specific points in $\mathbf{k}$-space as $T \rightarrow T_\mathrm{c}$ which abruptly change character into localized dips in the energy when the sample is cooled to $T < T_\mathrm{c}$. Note earlier NMR investigations on the double chain cuprate, YBa$_2$Cu$_4$O$_8$ by Suter et al \cite{Suter2000}, found evidence for unusual damping in the charge component of the NMR signals with a similar temperature dependence. This unusual damping can be explained by low frequency tails of the Giant Phonon Anomalies. In this way we have a consistent explanation for both the transition into the pseudogap state as observed in ARPES and high field Hall effect and also the unusual Giant Phonon anomalies in the pseudogap phase.

We have proposed a generalization of the well studied D-Mott insulator groundstate of the $\frac{1}{2}$-filled 2-leg Hubbard ladder to the 2D square lattice. The truncation of the Fermi surface in the ladder is caused by the presence of elastic umklapp scattering processes spanning the Fermi surface, which lead to a combination of strictly short range $d$-wave singlet pairing and commensurate antiferromagnetic correlations. Anderson's Resonant Valence Bond proposal shortly after the discovery of the cuprate superconductors, focussed on such spin singlet pair correlations as the origin of superconductivity in doped cuprates. Ossadnik's recent wavepacket formulation of many body theory is a crucial step in the investigation of nonperturbative modifications of electronic properties by SRO states in dimensions $D > 1$. 

Finally we would like to draw the reader's attention to a recent paper by Alexei Tsvelik \cite{Tsvelik2017} who has examined the relationship between the D-Mott insulator groundstate of the $\frac{1}{2}$-filled 2-leg Hubbard ladder and the 2D square lattice near the ``hot'' spots on the US within the strong coupling spin fermion model. Although his analysis is quite different from ours, the general conclusions are in agreement.

\acknowledgements This work was inspired by many illuminating discussions with Matthias Ossadnik on the wave packet description of the many body physics in short range ordered states. The authors are grateful to Alexei Tsvelik for insightful exchanges on the relationship between our sets of calculations. We also wish to thank Michele Dolfi, Manfred Sigrist, Peter Johnson, Matthias Troyer and Robert Konik for helpful discussions. Y.-H.L. is supported by ERC Advanced Grant SIMCOFE. Visits to Brookhaven Natl. Lab. by Y.-H.L. and T.M.R. are supported by the US DOE under contract number DE-AC02-98 CH 10886. W.-S.W. is supported by NSFC (under grant No. 11604168). Q.-H.W is supported by NSFC (under grant No. 11574134). F.-C.Z. is partly supported by NSFC grant 11674278 and National Basic Research Program of China (No. 2014CB921203).

\clearpage 
\appendix

\section{Details on DMRG calculation}

The DMRG is most efficient with open boundary conditions. There is however one subtlety related to the boundary effect in the presence of the $J$-term, as follows. At the boundary of the 2LL, a site has only 2 neighbors instead of 3 (as in the bulk). In this case a uniform $J$-term will attract particles to the boundary. If this happened, the calculated excitation energies would not represent bulk excitations in the thermodynamic limit. To overcome this problem, we replace the boundary rung interactions $J_{1,rung}=J_{L,rung}=2J$ as twice large, and keep everywhere else uniform. By inspecting the real-space particle-density distribution, we confirm the groundstate distribution is uniform and for excited states the extra particles are indeed doped into the bulk.

To include low energy excitations near the ladder Fermi points, we only choose $L$ such that $L+1$ is a multiple of 3, e.g. $L=14,32,62,95$. Because the groundstate of the Hubbard ladder at $\frac{1}{2}$-filling is unique and gapped, DMRG converges well (for the purpose of this paper) already at length $L=62$ and bond dimension $D=700$, as shown in Fig.~\ref{fig_WW}.

\section{Details on fitting ARPES}

In the 2D Hubbard model we have the kinetic parameters $t,t',t'',t'''$, which determine the Fermi surface. For each point on the noninteracting Fermi surface $\mathbf{k}_F$, we project it to $\mathbf{k}_\mathrm{U}(\mathbf{k}_F)$ on the US, and then calculate the projection of the velocity in the $(1,1)$-direction, $v=\left[ 
\partial_{k_x}\xi(\mathbf{k}_\mathrm{U})+ 
\partial_{k_y}\xi(\mathbf{k}_\mathrm{U}))\right]/\sqrt{2}$. Next we map the 4 $\mathbf{k}$-points generated by this point to the 4 Fermi points in a 2LL, whose hopping satisfies $2t_L\sin(2\pi/3)=v$. For simplicity we have chosen constant effective $U=0.4t$ and $J=0.1t$ for all sets of 4 $\mathbf{k}$-points. We use DMRG to get the $E_1$ in the ladder and then use it as $\Delta_R$ for the YRZ Green's function. Results are shown in Fig.~\ref{fig_ARPES}.

\end{document}